\documentclass[letterpaper,twocolumn,10pt]{article}
\usepackage{usenix2019_v3}
\usepackage[available]{usenixbadges}

\usepackage{tikz}
\usepackage{amsmath}

\usepackage{filecontents}

\usepackage{soul}
\usepackage{pifont}
\usepackage{xspace}
\usepackage{tabularx}
\usepackage{svg}
\usepackage{varwidth}
\usepackage{booktabs}
\usepackage{tabularx}
\usepackage{enumitem}
\usepackage{svg}
\usepackage{tikzsymbols}
\usepackage{subcaption}
\usepackage{multirow}
\usepackage{array}
\usepackage{hyperref}
\usepackage[group-separator={,}]{siunitx}
\hypersetup{
    allcolors=black
    }

\usepackage{colortbl}

\newcommand{\NUMVIDEOS}{\num{42986}\xspace}

\newcommand{\mdf}{MrDeepFakes\xspace}

\renewenvironment{quote}{%
  \list{}{%
    \leftmargin0.4cm   
    \rightmargin\leftmargin
  }
  \item\relax
}
{\endlist}

\newcommand{\quoteforum}[1]{%
\vspace{-2pt}
\begin{quote}\baselineskip=12pt%
    ``\textit{#1}''%
\end{quote}%
\vspace{-2pt}
}

\begin{document}

\date{}

\title{\Large \bf Characterizing the \mdf Sexual Deepfake Marketplace}

\author{
{\rm Catherine Han}\\
Stanford University
\and
{\rm Anne Li}\\
Stanford University
\and
{\rm Deepak Kumar}\\
UC San Diego
\and
{\rm Zakir Durumeric}\\
Stanford University
} 

\maketitle

\begin{abstract}
The prevalence of sexual deepfake material has exploded over the past several years. Attackers create and utilize deepfakes for many reasons: to seek sexual gratification, to harass and humiliate targets, or to exert power over an intimate partner. In part enabling this growth, several {markets} have emerged to support the buying and selling of sexual deepfake material. In this paper, we systematically characterize the most prominent and mainstream marketplace, \mdf. We analyze the marketplace economics, the targets of created media, and user discussions of how to create deepfakes, which we use to understand the current state-of-the-art in deepfake creation. Our work uncovers little enforcement of posted rules (e.g., limiting targeting to well-established celebrities), previously undocumented attacker motivations, and unexplored attacker tactics for acquiring resources to create sexual deepfakes.
\end{abstract}


\noindent
\begin{center}
    \fbox{\begin{minipage}{0.9\columnwidth}
    \textbf{Content Warning:}
        This paper includes content about image-based sexual abuse and other disturbing material.
    \end{minipage}}
\end{center}

\section{Introduction}
\label{sec:introduction}

The prevalence of deepfakes---media created or manipulated with artificial intelligence---has exploded over the last several years~\cite{deepfakes-everywhere-bloomberg}. Most deepfakes are nonconsensual sexual media: recent reports estimate that 98\% of online deepfakes are sexual in nature~\cite{homesecurityheroes2023state}. Beyond sexual gratification, attackers have created and disseminated sexual deepfakes to humiliate or harass targets\footnote{Drawing upon prior work on online harassment, we use ``target'' over ``victim'' to avoid disempowering those facing abuse.}~\cite{rana-ayyub-deepfake} or to exert control over intimate partners~\cite{lucas2022deepfakes}. These sexual deepfakes threaten the emotional, psychological, and physical well-being of targets~\cite{hao21horrifying}.\looseness=1

This growth is supported by dedicated markets that have emerged for buying and selling sexual deepfake content. These digital marketplaces are often accessible on the open Internet and enable buyers to commission custom sexual deepfakes from technically savvy sellers. The most prominent and mainstream platform is \mdf, a site where visitors can view celebrity sexual deepfakes, buy and sell deepfakes, and engage in community discussion on how to create deepfakes. Prior work has examined the community's migration away from Reddit, where it originated as \texttt{r/deepfakes} before being banned, to its new independent platform on {mrdeepfakes.com}~\cite{timmerman2023studying}. However, no work has studied this ecosystem's market dynamics, hosted content, or technical capabilities since the explosion of deepfakes post-2021.

In this paper, we systematically characterize the \mdf platform over the past five years.
We document the rise of the community and analyze the economics of its marketplace, focusing on the buyers, sellers, and targets of posted deepfakes. 
Combining comprehensive quantitative measurements of forum activity and video attributes with qualitative analysis of forum threads, we answer the following three research questions:
\begin{enumerate}[topsep=2pt,itemsep=0ex,partopsep=1ex,parsep=1ex]
    \item How has sexual deepfake video creation and consumption changed over time?
    \item What are the marketplace economics of buying and selling deepfake videos? 
    \item What are the requirements and challenges for creating deepfakes?
\end{enumerate}
\vspace{2pt}

As of November 2023, \mdf hosted 43K~sexual deepfake videos depicting 3.8K~individuals; these videos have been watched more than 1.5B~times. Analyzing the most common targets, who account for 95\% of videos on the site, we find that 95.3\% of individuals are women actors and musicians, making up 88\% of posted videos. 
While \mdf claims to only serve media depicting ``celebrity'' targets, we find these rules unenforced and that hundreds of targeted individuals have little to no online or public presence. Similarly, despite purported rules banning ``negative'' depictions of targets, we find that 2.4\% of uploaded videos include scenes of rape, abuse, or humiliation.





We next examine 830~paid request threads on the forum-based marketplace. Using thematic analysis, we surface three underlying motives for buyers: nonconsensual sexual gratification, power and masculinity, and community contribution. We identify the centering of community and the perception of commissioning deepfakes as contributing art
as a novel incentive that falls outside existing taxonomies of motives for online abuse~\cite{thomas2021hatesok,okolie2023ibsa}. 

While most request threads do not specify a price, we identify 58~posts that contain prices that buyers and sellers were willing to pay or accept. We find the average price of a deepfake video is \$87.50; prices ranged greatly and were as inexpensive as \$1 and as high as \$1,500. However, community guidance suggests that buyers should be wary of offers below \$50. Availability of facial image data for the target (``facesets''), output quality, difficulty of angles, and turnaround time are the primary factors that contribute to pricing.


Finally, we examine \num{1229}~posts across 25~deepfake creation guides and tutorial threads to study the current state-of-the-art. We document the three most crucial components that pose challenges for deepfake creators: access to facial image data, software proficiency, and powerful hardware. We find that the \mdf community relies on a host of resources, from open source software to free-tier cloud GPU access (e.g., Google Colab notebooks). We highlight that the community often references and utilizes recent academic research from well-established machine learning and computer vision venues and bypasses protections that developers have attempted to build into their tools to prevent abuse.


Ultimately, our work characterizes the sexual deepfake marketplace and documents the resources, challenges, and community-driven solutions that arise in the sexual deepfake creation process. We conclude with a discussion of our results, contextualizing implications of new incentives for online abuse, considerations for the open source and academic communities, and suggestions for 
how legislation may help.
By illuminating the largest public community behind creating and sharing deepfakes, we hope to provide valuable context for researchers, industry partners, and policymakers working to address this growing threat.
\section{Background \& Related Work}
\label{sec:related_work}

Our study is rooted in two bodies of related work: (1) interdisciplinary explorations of deepfake technology and the online communities that have formed around it, and (2) investigations of underground markets.


\subsection{Deepfakes and Image-Based Sexual Abuse}
\label{ssec:rel_deepfakes}




Literature across computer science, economics, and women's studies has sought to understand various aspects of deepfakes and other forms of image-based sexual abuse (IBSA), including online deepfake communities~\cite{timmerman2023studying,gamage2022deepfakes,widder2022ethicalai}, societal perceptions of deepfake consumption and production~\cite{brigham2024violation,mehta2023can,widder2022ethicalai},
and motives for perpetrating IBSA~\cite{okolie2023ibsa,kshetri2023economics,bellingcat2023any,thomas2021hatesok,scheuerman2021framework}.

Most directly related to our work, Timmerman et~al.\ investigated the emergence of the \mdf forum in response to the 2018 deplatforming of \texttt{r/deepfakes} from Reddit~\cite{timmerman2023studying}. 
Their work makes two primary contributions. First, they analyze how the community situates itself in society, such as its members' perceptions of the ethics, legality, and social acceptability of deepfakes. They show that deplatforming events have sowed distrust of mainstream platforms and the media within the community, positioning \mdf as the only viable source of technical support for creating sexual deepfakes.
Second, they characterize the targets of datasets shared within the community for use in deepfake creation, finding that they are primarily women celebrities from English-speaking countries.
Although our work also studies \mdf, we focus on analyzing the economics of its deepfake marketplace, the mushrooming of deepfake content, and the deepfake creation process (e.g., requisite software, hardware, data, and expertis).


Similarly, Widder et~al.\ found that core contributors and users of an unnamed, ``safe for work'' (SFW) open source deepfake tool believed that denying ``valuable technical support'' for usage in sexual contexts was their best means of control for its ethical use~\cite{widder2022ethicalai}.
In contrast, despite societal disapproval of creating and sharing deepfake nonconsensual intimate imagery (NCII)~\cite{brigham2024violation}, another study of deepfake discussion on Reddit concluded that online subcommunities are eager to facilitate both creating and sharing deepfakes~\cite{gamage2022deepfakes}. 

Despite past findings that depict deepfake creation as challenging, Mehta et~al.\ found that nonexperts can create deepfakes with relative ease; however, the poor realism of the participant-generated deepfakes indicates that creating photorealistic deepfakes remains difficult for novices~\cite{mehta2023can}. Still, while consumers of deepfake NCII often know it is fake and ``do not care,'' the harms experienced by their targets---emotional and psychological trauma, reputational damage, and financial loss~\cite{dickson2020tiktok, ayyub2018silence, dhsincreasing}---are independent of deepfake content's achieved realism~\cite{burkell2019nothing}. 

Beyond deepfakes, extensive prior work has sought to understand online abuse---the underlying motives of perpetrators, targets' experiences of harm, and their risk factors---producing several frameworks and taxonomies~\cite{thomas2021hatesok,warford2022framework,scheuerman2021framework, marwick2021networked,phillips2019exclusionary,han2023hate}.
For deepfake NCII and other forms of IBSA, studies have identified motives including power, circumvention of consent, revenge, and financial gain~\cite{okolie2023ibsa,thomas2021hatesok,bellingcat2023any,kshetri2023economics}.
Using an economics framing, Kshetri introduced a cost-benefit model for abusive deepfake creation that accounts for investment, opportunity, and psychological costs and financial and psychological benefits~\cite{kshetri2023economics}.
We draw on these frameworks to interpret deepfake NCII using a scoped threat model~\cite{thomas2021hatesok} with three stakeholders: the \textit{consumer}, the \textit{creator}, and the \textit{target}.
We define the target as one whose likeness is abused out of context; however, sex workers whose bodies nonconsensually appear in these videos are also harmed, and they remain a stakeholder population to be studied in future work.

\subsection{Underground Forums and Marketplaces}
\label{ssec:rel_underground}


The security community has previously studied online underground forums and markets, including the purveyance of stolen credentials and drugs~\cite{motoyama11analysis, soska15measuring}, bulletproof hosting services~\cite{noroozian19platforms}, and ransomware~\cite{gray2022money}. The profit-driven framing of attackers in such ecosystems has allowed researchers to model them as rational actors, using economic principles to identify potential disruption tactics.
Tseng et~al. applied similar methods to partner infidelity and surveillance forums, examining how members carry out intimate partner surveillance~\cite{tseng20tools}. They observe that these communities
openly discuss and develop new abuse strategies in public forums, concluding that improving moderation has mitigation potential~\cite{tseng20tools}. 

Researchers have also identified vulnerabilities in the monetization infrastructure of cybercrime operations (e.g., sex trafficking, ransomware, spam) to leverage for the detection, attribution, or disincentivization of such activity. These potential disruption channels include Bitcoin transactions~\cite{portnoff17backpage}, cryptocurrency exchanges~\cite{gray2022money}, and payment processing and online advertising providers~\cite{bellingcat2023any,han2022infra,levchenko2011trajectories}. 
This body of work demonstrates the promise of mitigations informed by measurements of underground forums and markets; we employ similar approaches to study the deepfake NCII ecosystem.


\begin{figure}[h]
   \centering
   \includegraphics[width=0.8\linewidth]{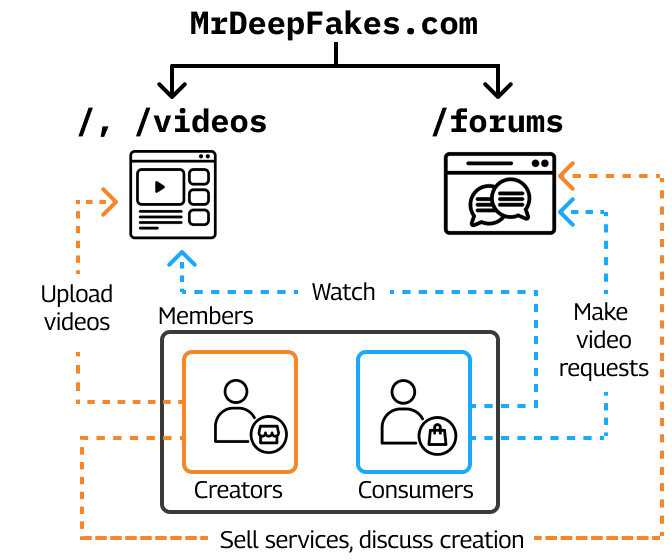}
   \caption{\textbf{\mdf}---The \mdf platform is composed of two spaces: a tube site for watching sexual deepfake content and a forum where buyers can solicit content.}
   \label{fig:mdf_platform}
   \vspace{-10pt}
\end{figure}

\subsection{\mdf Platform}
\label{ssec:bg_platform}
MrDeepFakes is the largest open sexual deepfake community~\cite{meikle2022deepfakes, maiberg22clicks} and consists of  
two spaces: (1)~a media sharing platform where members upload deepfakes, and (2)~a forum for discussing deepfake-related topics (e.g., how to create deepfakes) and negotiate the exchange of deepfake-related commodities (e.g., videos and datasets) (Figure~\ref{fig:mdf_platform}). The former is served as the landing page of \mdf as a ``tube'' site\footnote{A pornographic video-sharing site (e.g., PornHub)} where visitors can watch deepfake videos. 








\section{Data Collection}
\label{sec:data_collection}

Our study is based on crawling the \mdf website using headless Chrome and the Selenium WebDriver. We specifically collected public forum posts and replies, metadata attached to public videos, and public user profile information in October and November of 2023. We created an account to access public site data, but we did not bypass any security controls, attempt to access any private data, download any video or image content, or interact with forum members. We never attempt to deanonymize or investigate individual buyers or sellers. We describe the ethical implications of our data collection and analysis in greater detail in Section~\ref{subsec:data_collection_ethics}.

\vspace{3pt}
\noindent
\textbf{Deepfake Videos.}\quad
Starting at \texttt{PAGE=1}, we paginated over \texttt{/videos/[PAGE]} until we hit an HTTP 404 \texttt{Not Found} error. This yielded URLs for \NUMVIDEOS~videos that we crawled to collect each video's title, description, creator, measures of engagement, tagged individuals (deepfaked celebrities), and any provided Bitcoin address for donations to the uploader. 

\vspace{3pt}
\noindent
\textbf{Forum Posts.}\quad
We scraped each forum section landing page and collected URLs to each subsection and thread, which we scraped to collect the title and body along with associated responses, including message timestamp, author, and attachments. In total, we scraped 8,198~threads, comprising 43,350~forum posts. Table~\ref{table:section_thread_count} shows a breakdown of each crawled forum section, and we provide the hierarchical section structure in Appendix~\ref{app:forum_map}.

\begin{table}[t]
\centering
\small
\begin{tabular}{lrr}
\toprule
 Section   & Threads  &  \% \\ \midrule
DeepFake Video Creation Tools     & 4,666                                    & 56.92                                  \\
Adult DeepFake Videos (NSFW/18+) & 1,593                                    & 18.43                                  \\
Image Fakes (NSFW/18+)           & 714                                      & 8.71                                   \\
Lounge                                                                       & 590                                      & 7.20                                   \\
Non-Adult Content (SFW/All Ages) & 464                                      & 5.66                                   \\
Stable Diffusion                                                             & 82                                       & 1.00                                   \\ 
Other Languages                                                              & 48                                       & 0.59                                   \\ 
Marketplace                                                                  & 26                                       & 0.32                                   \\
Announcements                                                                & 15                                       & 0.18                                   \\
\midrule
\emph{Total} & 8,198 & \\

\bottomrule

\end{tabular}
\vspace{-4pt}
\caption{\textbf{Forum Sections}---We present the breakdown of activity by thread count in each \mdf forum section.}
\label{table:section_thread_count}
\vspace{-5pt}
\end{table}

\vspace{3pt}
\noindent
\textbf{User Accounts.}\quad
Opening a \mdf account creates two public profiles by default: one on the forum and another on the tube site. On the forum, users are assigned monotonic user IDs, which we crawled by visiting \texttt{/forums/members/[USER\_ID]} up to the highest known user ID (the forum links the profile of the newest user at any given time on its homepage). For each profile, we scraped the ID, banners (which denote special roles like moderator), profile posts, and bio. We collected data about 611K~public profiles. 

The tube site's schema for assigning user IDs is independent of the forum (i.e., they do not match and are not monotonic). To scrape its public member profiles, we visited pages with the path \texttt{/users/[USERNAME]}, based on forum usernames. 
%
%
As with videos, we collected profile URLs, then scraped data about each profile, including subscriber count, biographical information, and Bitcoin donation address. We collected 595K~entries, an upper bound on the number of profiles, since some entries served a custom ``not found'' page without an explicit 404 response code. 

\label{subsec:data_collection_limitations}

\subsection{Limitations}
\label{ssec:limitations}

Our analysis is limited to publicly available data, which does not capture private activity on the site or communication on third-party platforms; we also exclude non-English discourser work likely underestimates the degree of forum activity and does not capture more fringe and illicit behaviorshurn in forum, video, and user profile content also prevented some data from being collected. We validated the completeness of our data by comparing it to the content on the website as rendered in our browser shortly after our data collection, and we manually verified the correctness of 10\% samples of forum, video, and user data.
\begin{figure*}
    \centering
    \begin{subfigure}[t]{0.33\textwidth}
        \centering
        \includegraphics[width=\linewidth]{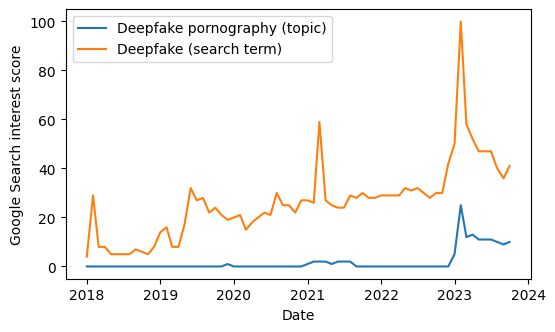}
        \caption{Internet Interest}
        \label{subfig:search_trends}
    \end{subfigure}%
    \begin{subfigure}[t]{0.32\textwidth}
        \centering
        \includegraphics[width=\linewidth]{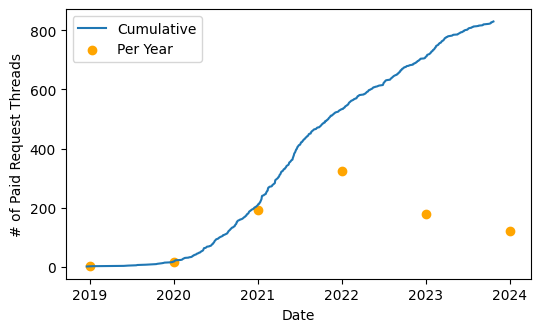}
        \caption{Paid Requests}
        \label{subfig:num_paid_reqs}
    \end{subfigure}%
        \begin{subfigure}[t]{0.33\textwidth}
        \centering
        \includegraphics[width=\linewidth]{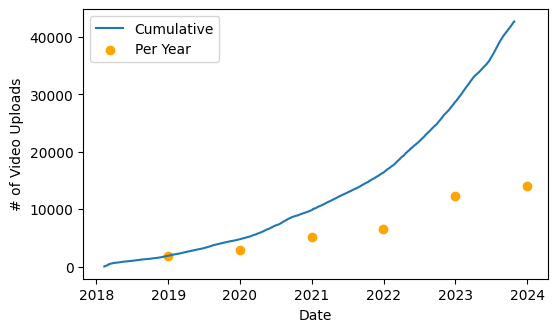}
        \caption{Posted Videos}
        \label{subfig:num_vid_uploads}
    \end{subfigure}%
    \caption{\textbf{Deepfake Interest and \mdf Growth}---Google Search interest score in deepfakes and deepfake ``pornography'' has increased significantly since early 2021, coinciding with growth in paid request and video upload activity on \mdf.}
    \label{fig:mdf_growth}
    \centering
    \vspace{-5pt}
\end{figure*}

\section{\mdf Tube Site}
\label{sec:characterizing}

Founded in 2018, \mdf experienced relatively little engagement during its first few years online~\cite{timmerman2023studying}. However, following a broader spike in sexual deepfake interest starting in 2021 (Figure~\ref{subfig:search_trends}), \mdf' popularity has grown rapidly. The site homepage's interface emphasizes watching uploaded sexual deepfake videos, and as of July 2024, \mdf is one of the 10K~most popular websites globally according to Google CrUX~\cite{durumeric2023cached, ruth2022toppling}. At the time of our data collection in November 2023, the site hosted \NUMVIDEOS~sexual deepfake videos posted by 1,880~users, which garnered over 1.5B~views and a median 14.3K~views per video. In this section, we analyze the individuals targeted by deepfake videos uploaded to \mdf.

\subsection{Methodology}

To understand the individuals targeted by deepfakes, we analyzed videos uploaded to the \mdf tube site. Posted deepfake videos have attached metadata that enables discoverability on the site, including the name of the depicted (i.e., deepfaked) target, descriptive categories (Table~\ref{tbl:top_categories}), and free-form tags. Over 90\% of videos specified the depicted celebrities, which enabled us to identify and analyze the demographics of 3,803~unique individuals depicted in deepfake videos. 

\begin{table}
    \centering
    \begin{minipage}{.4\columnwidth}
        \small
        \setlength{\tabcolsep}{5pt}
        \begin{tabular}{lrr}
        \toprule
        Occupation & \multicolumn{2}{c}{Frequency} \\ \midrule
        Actor & 1,130  & 58\% \\
        Musician & 578 & 30\% \\
        Model & 336 & 17\% \\
        TV & 245 & 13\% \\
        Content Creator & 231 & 12\% \\
        \bottomrule
        \end{tabular}
    \end{minipage}
    \hspace{28pt}
    \begin{minipage}{.4\columnwidth}
        \small
        \setlength{\tabcolsep}{5pt}
        \begin{tabular}{lrr}
        \toprule
        Occupation & \multicolumn{2}{c}{Frequency} \\ \midrule
        Personality & 160 & 8\% \\
        Filmmaker & 82 & 4\% \\
        Sex Worker & 65 & 3\% \\
        Writer & 63 & 3\% \\
        Reporter & 60 & 3\% \\
        \bottomrule
        \end{tabular}
    \end{minipage} 
    \vspace{-4pt}
    \caption{\textbf{Top Occupations of Targets}---We categorize the occupations of the most commonly depicted individuals in posted deepfake videos, who account for 95\% of videos. Labels are not mutually exclusive.}
    \label{table:occupation}
\end{table}

\begin{table}[h]
\centering
\small
\begin{tabularx}{\linewidth}{ X r r r }
\toprule
Category                & Videos & Total Views & Median Views \\
\midrule
Hardcore                & 15.6K     & 763.9M          & 20.6K        \\
Blowjob                 & 15K       & 628.5M          & 16.5K        \\
Asian                   & 14K       & 354.8M          & 10.6K        \\
Korean                  & 7.6K      & 137.5M          & 9.3K         \\
Cumshot                 & 7K        & 347.3M          & 19.2K        \\
Facial                  & 3.6K      & 192.1M          & 22.3K        \\
Indian/Bollywood        & 2.9K      & 291.4M          & 35.1K        \\
Interracial             & 2.4K      & 107.0M          & 22.5K        \\
Porn Music Videos       & 327       & 123.4M          & 23.0K        \\
AI Voice                & 36        & 1.2M            & 20.9K        \\
\bottomrule
\end{tabularx}
\vspace{-5pt}
\caption{
    \textbf{Top Video Categories}---The most popular deepfake video categories, sorted by the number of uploaded videos. Categories are not mutually exclusive.
    }
\vspace{-10pt}
\label{tbl:top_categories}
\end{table}

\subsection{Targets of Deepfake Videos}

Targeted individuals follow a long-tailed distribution: the ten most targeted celebrities are depicted in 400--900~videos (1--2\% of all videos) each, but over 40\% of targets are depicted in only one video. We manually investigated the \num{1942} celebrities most commonly depicted, who account for over 95\% of videos, on the basis of gender/sexual identity, occupation, nationality, and social media following using Google search, news stories, celebrity websites, and social media profiles.

The 20~most targeted individuals are well-known women actresses, or popular American or Korean musicians.
Just over one third of targeted individuals are American, followed by Koreans (11\%), many of whom are Korean pop (``K-pop'') celebrities. MrDeepFakes' focus on sexual rather than political disinformation deepfake content appears to minimize the number of depicted geopolitical figures (43~individuals,~2.21\%). 

Consistent with industry reports~\cite{homesecurityheroes2023state, deeptrace2019state}, we find that 91\% of \emph{videos} target gender-marginalized individuals (i.e., anyone whose labels do not consist solely of ``man'' or ``unknown,'' including women and LGBTQIA+ people). Of the \num{1942} \emph{targeted individuals}, 95.3\% are women, and 2.1\% identify as LGBTQIA+\@. Most targeted individuals are actors, musicians, television personalities, or social media influencers (Table~\ref{table:occupation}). Actors comprise more than half of the targets (58.19\%), possibly in part because more high-quality and readily available facial data exists in the form of public images and videos. 

\subsection{Non-Celebrity Targets}
\label{subsec:characterization_nonceleb}

MrDeepFakes advertises itself as a ``celebrity deepfake porn tube site,'' and its community codifies permitting only sexual deepfakes that depict celebrities. Website rules state, among other requirements, that a celebrity must have ``significant influence'' on social media with  over 200K~followers on TikTok or 100K~followers on Instagram, YouTube, Twitter, or Twitch. 
Despite these rules, 271~of the 1,942~individuals we investigated (14\%) fall far below these thresholds. For 29~targets, we found no online presence, and for another 242~targets, we could not find profiles on any of the listed social media platforms that satisfied the minimum following criteria. In one example, we find that 38~Guatemalan newscasters with little to no social media following appear in over 300~videos. All of these videos were posted by two users, who both describe specializing in Latin American individuals in their profiles. 

These statistics likely \emph{significantly underestimate} how often non-celebrities are targeted in sexual deepfake videos because our investigation is limited to the top 50\% of the most commonly depicted individuals in public videos. At the same time, we acknowledge the possibility of targets deleting social media accounts after, or perhaps due to, being depicted in deepfake content. Therefore, our measurements may also potentially overestimate non-celebrity targets.

\subsection{Assault- and Abuse-Based Content}

\mdf rules prohibit posting of content that ``depicts celebrities in a negative manner.''
While ``negative manner,'' is not defined, they list ``rape scenes, abuse, [and] humiliation'' as prohibited examples, noting that users who post such content will receive a warning if the source video is ``professional porn'' and will otherwise be banned. Despite this, searches for terms related to violence, assault, rape, abuse, humiliation, and ``gang bang'' 
yield 1,017~videos (2.37\%). Some depict the targeted individual as the perpetrator, rather than the victim, of such abuse, extending beyond nonconsensually sexualizing targets to creating slanderous and criminal imagery. 

\vspace{5pt}
\newpage
\noindent
According to the site footer, \mdf' stated purpose is to ``provide a safe-haven without censorship'' for people to experiment with deepfake technology, within the rules of the community. While this ``safe-haven'' has welcomed a substantial influx of new members, our findings indicate that deepfake creators are undeterred from generating harmful and inherently nonconsensual content and publicly distributing it on \mdf. Such behaviors violate even the code of ethics expressly defined by their own community.

\section{\mdf Marketplace}
\label{sec:market}
In addition to its tube site, \mdf hosts a forum-based {marketplace} for members to request free or paid sexual deepfake videos. Aligned with the site's overall growth, public requests for deepfake videos more than tripled since 2021 (Figure~\ref{subfig:num_paid_reqs}). While there are also forum sections dedicated to the paid requests of celebrity and sex worker image datasets and trained models, they appear unused, with only three threads requesting celebrity datasets and none in the other two. In this section, we analyze requests for deepfake videos and their responses to understand the content sought by buyers and the costs associated with purchasing custom sexual deepfake content.

\subsection{Methodology} 

We perform thematic analysis~\cite{braun2012thematic,kiger2020thematic} for a random subset of 263~(of 830)~paid requests for deepfake videos. Our initial codes were guided by primary topics of interest from manual exploration of the marketplace and related work (e.g., buyer motivations grounded in those codified by Okolie et al.~\cite{okolie2023ibsa} for IBSA perpetrators, buyers' methods for vetting sellers, desirable attributes of deepfake videos). Codes were iteratively added to the codebook (Appendix~\ref{app:codebook}) as we processed our dataset and adjusted recursively as our process surfaced themes (e.g., the degree to which buyers and sellers positioned themselves as artistic contributors). Two researchers independently assigned codes to the posts, and codes were not mutually exclusive. Inter-rater reliability (IRR) was reasonably high with a Kupper-Hafner metric~\cite{kupperhafner1989irr} of~0.72. For posts where the set of assigned codes were not identical, the researchers met to discuss the conflicts and reach agreement to obtain a final coded dataset. Presented quotes were jointly paraphrased by two authors.

\subsection{Requests for Deepfakes}
\label{ssec:consumers}

The \mdf marketplace is request-based: buyers request free or commission paid deepfake videos in the forum section \emph{Adult DeepFake Videos (NSFW/18+) / Video Requests}.  
As of November 2023, there were 830 (66.9\%)~paid and 411 (33.1\%)~unpaid requests. 94.7\% of requests were responded to: paid requests received an average 2.6 replies, while unpaid requests received an average 4.6~replies. Over 90\% of the 707~unique buyers made exactly one public request; this may be due to later transactions occurring privately once a buyer has identified a satisfactory seller.


The vast majority (96.2\%) of paid requests on \mdf are for ``adult deepfake'' videos, as designated by its parent section (Appendix~\ref{app:forum_map}). However, we also find requests for non-sexual or ``SFW'' deepfakes (4), machine learning models for creating deepfakes~(3), and data used in deepfake creation~(2).
Seven buyers specified that making a deepfake request is what drove them to join \mdf or to post for the first time; 13~buyers requested multiple videos or promised longer-term patronage if satisfied with the finished product.

Requests provided a range of details for technical and target specification. 10.9\% of requests are intentionally vague about the identity of their target. On the other handmany, m requests not only specify their target, but they also discuss the availability of targets' facial image data (e.g., linking to targets' social media accounts, listing movies featuring the target). Further, 31~posts (11.7\%) advertise that they can facilitate the creation process by providing creators access to private or personally-curated repositories of source or destination image data (e.g., high-resolution video files, personal videos taken of musician targets at shows).


\subsubsection{Buyer Motivations}
\label{sssec:consumer_motives}

Prior work has predominantly described the motives for deepfake abuse as either seeking nonconsensual sexual gratification or producing political disinformation~\cite{okolie2023ibsa,thomas2021hatesok,dhsincreasing}. While we do find sexual gratification to be a major motivator, we find others as well.\looseness=-1

\vspace{3pt}
\noindent
\textbf{Nonconsensual Sexual Gratification.}\quad
Nearly a quarter (23.7\%) of requests are explicitly connected to the buyer's pleasure (e.g., sexual remarks about the target in the request post). Several posts specifically seek extreme (i.e., racialized or violent) or fetish content.\footnote{We use the term ``fetish'' as defined in DSM-5, where someone's sexual pleasure depends on nonliving objects (such as undergarments or high-heeled shoes) or a highly specific focus on a body part (most often nongenital)~\cite{kafka2010fetish}.} The majority of the targets on \mdf are not personal acquaintances of the buyers or sellers of the content that depicts them. Therefore, the buyers' motives are not captured by understandings of other IBSA perpetrators, such as producers of ``revenge porn.'' However, we do find instances of requests that demonstrate the \emph{parasocial nature} of deepfake NCII. For example, in two requests for content depicting an actress and a social media influencer, buyers expressed perceived intimacy with their targets (e.g., referring to the target as ``their'' target, claiming that they not only ``love'' her, but that they love her ``the way she is'').

\vspace{3pt}
\noindent
\textbf{Power and Masculinity.}\quad
We find anecdotally that some buyers seek to weaponize the creation of deepfake NCII to feel empowered or exert control---physically, emotionally, and/or financially---over their target:
\quoteforum{
    I WANT HER DONE FOR 2 REASONS: 1. she's hot \\ 2. she has a huge ego and maybe this will humble her.
}\vspace{-5pt}
\quoteforum{
     Now that she's doing porn but her looks are gone, I bet she'd like having her younger self's face used for hard porn. If you do it, consider giving her a cut of the profits---she probably needs the cash.
}

\noindent Likewise, one buyer clearly expresses violent attitudes toward women, stating that the creator would get ``bonus points'' if they can make it ``rough'' or ``painful,'' tying the buyer's sexual gratification to violence against women. Another request infantilizes its target as a ``daddy's daughter,'' using her depiction in sexual scenes to fantasize about physically punishing her.
All of these buyers received replies from interested sellers, and
none were reprimanded about the kind of content they requested.
Psychology literature has investigated the norms of masculinity, finding that controlling and subordinating women are significant indicators of conformity to masculine norms~\cite{mahalik2003cmni}.
The normalization of buyers' hyper-masculine expressions and preferences for sexual content by other buyers and sellers demonstrates \mdf market participants' conformity to prototypical masculinity.

\vspace{3pt}
\noindent
\textbf{Community Contribution.}\quad
In 7\% of requests, the buyer situates their request as benefiting the community, or even more broadly, consumers of pornography. The most common manifestation of this sentiment is the buyer expressing that they (1)~failed to find content fitting their parameters on the site and (2)~are seeking to fill this ``gap,'' believing it would benefit the community, whose members share values and sexual preferences with the buyer. For example, one request appeals to potential creators by stating that their request would be a good fit for members of \mdf:

\quoteforum{
    I know TONS of folks want deepfakes of [redacted]\ldots I promise the video would be very popular on \mdf and more generally, she hardly exists outside of Instagram and Twitch.
}


\noindent
We also find evidence across 10 requests that deepfake consumers esteem their requests as \textit{artistic} pursuits and perceive deepfake creators as talented \emph{artists}:
\quoteforum{
    It would be an honor to commission the first masterpiece of [redacted]'s debut into the world of hardcore porn\ldots I can't afford an extravagant price so I hope to connect with an artist that shares my vision.
}
\vspace{-5pt}
\quoteforum{
    Listen---it might seem odd to request a deepfake of pornstars, but for porn connoisseurs like me, we know that [redacted\_1] has not done a BBC scene, but her lookalike [redacted\_2] has\ldots
}

\noindent The security community has taxonomized the harms of online abuse, characterizing perpetrators as driven by the desire to inflict physical, psychological, or sexual harm to, silence, or coerce targets~\cite{thomas2021hatesok}. However, the perception of deepfakes as \emph{art} and of their consumers as \emph{connoisseurs} introduces a new intent, which we discuss in Section~\ref{ssec:disc_community}. We study the deepfake creation process and how the \mdf community supports inexperienced creators in Section~\ref{sec:pipeline}.  


\subsubsection{Requests for Non-Celebrity Deepfakes}

Despite forum rules prohibiting requests for non-celebrity targets and requiring inclusion of target celebrities' names (``[n]on-celebs'' cannot be requested), 11\% of requests did not identify their target. Furthermore, eight of these requests specified that the targets were \textit{not} celebrities. Two in particular described targets from the requesters' personal lives: (1) a girl that the buyer ``liked'' and (2) the buyer's partner. 

These requests were not taken down or admonished by moderators, as we observed in another request thread where the buyer claimed to want a deepfake of \textit{themselves} and would ``provide a (sic) consent for sure.'' In this thread, a moderator, who conceded that \mdf did not have rules against ``requesting deepfakes of yourself,'' warned the buyer:
\quoteforum{
     Some creators will require verification that it's you before they start. Also, consider how much you trust the creator because there are no safeguards to keep your video from being uploaded to the public Internet (not our forum, since we moderate any non-celeb uploads).
}
Further, the moderator specified only that \textit{some} creators would require a clear demonstration of consent for such requests, which implies a range of ethics among deepfake creators and who they consider permissible to deepfake. Ultimately, the two requests for non-celebrity deepfakes received responses from interested sellers who were verified creators on \mdf. The request for what is alleged to be the buyer's partner appeared to be fulfilled:
\quoteforum{
    Request fulfilled, thanks so much. Got the best quality deepfake within a week after providing just a few photos. I would highly recommend him. He even provides a sample early in the process and incorporated my feedback into the final masterpiece.
}

\noindent The ability of buyers to find less scrupulous sellers parallels what prior investigations of underground markets have documented with alternative service providers who are tolerant of criminal or abusive behaviors~\cite{noroozian19platforms,kanich2008spamalytics,levchenko2011trajectories,motoyama11analysis,franklin2007inquiry,han2022infra}. The ambiguous and informal processes for verifying the alleged consent of private targets raise further concerns: buyers can abuse these gaps to thinly veil nonconsensual deepfake requests for \emph{any} target as consensual and find willing deepfake creators. It is in these cases that we may anticipate the abuse of the paid requests marketplace to create deepfakes of personal acquaintances or enemies of the buyer, which more closely resembles the motives underlying other forms of IBSA, such as sextortion, revenge, and reputational harm.

\subsubsection{Requests for Deepfakes Targeting Children}

\mdf prohibits deepfakes that target underage individuals. The site's rules also note a blocklist of individuals who have only recently turned 18~years old and should not be targeted due to uncertain legality of available image data, but as of August 2024, the list is empty. One buyer in our dataset appears to request content sexualizing children:
\quoteforum{
    I need 1 image of 2 boys by today. The swapped final picture will be 18+ with me in it. Contact me quickly at email address [redacted].
}

\noindent
While this thread did not receive any replies, it was also not taken down or given a warning by a site moderator, like many other posts on the site. It is unclear whether the request was fulfilled through private communication. 

\subsection{Creators and Sellers}
\label{ssec:sellers}

Over 1,800~users have uploaded deepfake video content on \mdf. We note that \num{657}~users (0.1\%) uploaded over 95\% of the site's video content, in line with the distribution of user participation on other social media platforms~\cite{nielsen2006participation}.  Around 228~users have responded to requests for free or paid deepfake videos, of which 55~are ``verified'' sellers (Section~\ref{ssec:verified}); however, this likely underestimates the total number of users creating videos, since a significant fraction of communication occurs over private channels. 94.7\% of paid and 72.9\% of free requests are replied to, and the median times to receive a response are 16~hours and 5~days, respectively.

Sellers responded to an average eight posts, but this is heavily skewed with a handful (10--15) of sellers responding to an outsized number of requests. One seller responded to 531~requests alone. In these cases, sellers appear to aggressively respond to most posts with generic replies like ``PM sent,'' ``Please check your PM, thank you,'' and ``Here is a sample of my work: [...] dm me on discord: [...]'' 

\subsubsection{Marketing and Differentiation}
\label{sssec:monetization_tech}

Sellers commonly showcase their abilities to potential buyers by sharing their own deepfake videos on \mdf' tube site. In paid request threads, 365 (18.9\%) of deepfake sellers' replies included links to deepfake videos on the tube site as well as embedded screenshots or thumbnails from past work.
Because deepfake approaches are rapidly evolving and creators want to showcase their best work, this also appears to lead to churn in hosted videos. We observed that \num{1948} of \NUMVIDEOS video URLs were inaccessible; 75.1\% of deleted videos indicated that they had been deleted upon creator request. Anecdotally, we find increasing specialization among deepfake creators, like we observed with creators dedicated to deepfaking Guatemalan newscasters (Section~\ref{subsec:characterization_nonceleb}). In response to a request seeking ``somebody who would specialize [in] classic (1970s and 1980s)'' deepfakes, a verified creator, moderator, and staff member responded:
\quoteforum{
    I definitely would, but my specialty is more placing today's celebrities into 70s porn\ldots If you suggest a celeb that inspires me, I might do it.
}
\vspace{-2pt}

\vspace{-5pt}
\subsubsection{Verified Creators}
\label{ssec:verified}



Deepfake creators can apply to become \emph{Verified Video Creators}, who receive a ``trusted'' badge indicating that the user creates their own deepfake content and is able to monetize on the platform (e.g., paid video downloads, cryptocurrency payouts, forum advertising privileges). Verified creators also bypass content moderation, which site administrators claim is a process that all unverified users' uploads undergo and takes several days. Further, platform affordances incentivize verification: the tube site showcases verified creators' videos, and a verified creator's verification status appears as a colorful label alongside their forum posts. The application process aims to certify that they generate their own content. To demonstrate their required technical knowledge for deepfake creation, applicants must provide evidence that they possess datasets or images of applications they use, and they must also describe their creation process. Beyond proving their capabilities, the applicant must also have reasonable standing within the \mdf community, as they are required to link to their tube site profile with at least three ``active'' videos (i.e., not removed by moderators). Their account must also be e-mail verified, at least 30~days old, and cannot have received warnings from moderators. 


The verification process appears to be somewhat selective; only 488~users are verified video creators, which is less than a third of all users who have uploaded to the tube site. Of the verified users, 55 (11.3\%) have replied to paid requests.
In addition to verification, there are other designations of authority and reputation, or ``banners,'' such as staff member, administrator, moderator, and developer. While banners are not mutually exclusive, 96.4\% of these users only have the verified video creator banner; there are only four users that have more than one banner. 
Of the verified creators selling their services, two also hold positions of power within the community: both are staff members and moderators.

\vspace{-1pt}
\subsection{Cost of a Custom Deepfake}
\label{ssec:cost}

Most price negotiations occur privately, and 15\% of initial requests explicitly ask interested creators to direct message them on the forum or through an off-platform channel (e.g., e-mail or Telegram) instead of replying publicly. However, we find that 11~requests (4.2\%) in our sample included concrete prices that the buyer would pay. We searched for currency symbols across the 2,863~paid requests and replies, which surfaced 184~posts (6.4\%) discussing costs. After removing spam and other irrelevant posts, we identified 58~posts that included prices that buyers were willing to pay for deepfake videos and/or that sellers were willing to accept.

The average price advertised by sellers is \$87.50; 35.1\% of quoted prices are at least \$100, and 12.3\% are at least \$200. The range in prices is significant: the least expensive advertised service was \$1 for a one-minute video with a 30-minute turnaround time with one condition: ``No [blowjobs].'' Conversely, the highest price discussed was \$\num{1500}. For 31 of the 58~request posts with explicit prices, the length of the video was included as a parameter; the length of a deepfake video is loosely correlated with the commission price ($R = 0.29$). From the limited visibility of a public vantage point, we cannot meaningfully determine if prices have changed over time. However, we do find anecdotes from community leaders that describe how deepfake technology has shifted the economy of the broader and longer-standing ``faking'' market:
\vspace{-10pt}
\quoteforum{
    Manual faking might have been killed by AI, but not faking in general. AI has made faking more accessible and now, it's even harder to make money from it than before (it was already hard because of laws and policies on platforms like Patreon). Still, AI has sped up the faking process and lowered the barrier to entry for folks with less time or technical skills to make money off fakes (you aren't limited to celeb fakes anymore, especially with tools like DeepFaceLab and Roop).
}

\noindent This suggests that the accessibility of deepfake creation, relative to that of manual fakes, has shifted the economy to meet a growing demand for \textit{non-celebrity} deepfakes. We examine the deepfake creation process in Section~\ref{sec:pipeline} and the implications for deepfake NCII in Section~\ref{sec:discussion}.


\subsubsection{Factors Affecting Price}

Several buyers noted the factors that affected the price that they were willing to pay; they were willing to pay more for videos with more than one target depicted, if the angles of the video were complicated, or if their target was male. Because predominantly women are targeted by deepfake NCII, facesets and information regarding sex worker ``body matches'' for men are comparatively rarer. Several buyers described the factors they considered when choosing a seller to fulfill their request. Most commonly (12\% of sampled requests), buyers explicitly stated that they sought high-quality deepfakes where a deepfake's quality is typically defined by its resolution or the degree of achieved realism. Other factors included the estimated production turnaround time (1.9\%) and the ``professionalism'' of the seller (0.8\%). To proactively vet sellers, 12~requests asked interested creators to link to a portfolio of their past work; two other requests specified they were looking for a verified video creator (Section~\ref{ssec:verified}) or one with good reviews on the dedicated forum thread.

Inexpensive services (under $\approx$\$50) are described by moderators and other creators as being poor quality and not worth buyers' money. A moderator of \mdf, who also advertises their services and develops forum guides for deepfake creation, warns buyers of such offers, or ``cheapfakes'':
\quoteforum{
    Beware of cheap offers for \$10--50 fakes, they can't possibly deliver quality. One guy who frequently advertises their services here did such a shit job for one of my clients that I would have banned him.
}

\noindent The overall market for deepfakes has thus split into two tiers: (1) low-quality and fast ``cheapfakes'' and (2) high-end, custom ``art.'' The emergence of commodity deepfake applications (e.g., those marketed for face swapping, undressing, nudifying, etc. capabilities)~\cite{westerlund2019deepfake,cbs2019cheapfake,paris2019cheapfake} caters to the former, while seller scrutiny of cheapfakes situates the paid requests marketplace on \mdf as offering the latter. These higher-end figures highlight the technical challenges posed in deepfake creation, which appear to require the expertise of a highly-skilled professional or the money to pay one. 

The condition attached to the \$1 offer also hints at the technical challenge posed by deepfaking scenes of oral sex due to facial obstructions and complicated angles, which is widely discussed across the forum. A brand account for a commodity deepfake software, Swapface, admits in a thread where \mdf members were constructively critiquing its outputs that, ``partial face and side face is really a big challenge for all deepfake methods'' that they would attempt to address in the next update. Similarly, the \$\num{1500} figure was suggested by another \mdf moderator and deepfake creator regarding a commission for a movie scene so complicated that they declared it impossible via AI methods and would instead require more ``traditional,'' manual methods:
\quoteforum{
    The only answer is to find a legit SFX guy working in TV and film, pay him \$500 to \$1,500 at least. It's probably even more to match the lighting and body parts nicely, it'll take at least 2 weeks of work even for a professional.
}
Another moderator agreed with this evaluation, advising the buyer to either ``prepare plenty of [B]itcoins'' or choose a simpler scene, implying that such a difficult request would likely be prohibitively expensive and require niche expertise. 


We find that sellers potentially establish private communication channels with potential buyers even before the price negotiation step. Of the \num{32284} videos that have non-null descriptions, using regex and keyword search, we identify that 24.9\% of the video descriptions provide viewers with contact information or otherwise solicit private communication. This positions videos uploaded to the tube site as advertisements themselves. Offers to communicate via first-party direct messages on the \mdf site are found in 8.9\% of videos. Third-party platforms are also popular; the most common third-party platforms for direct messages include Telegram, Discord, and e-mail. We note limited use of third-party platforms that offer end-to-end encryption (E2EE) channels, identifying three references to Proton Mail addresses and 21 references to WhatsApp contact information. While Telegram offers E2EE in its ``Secret Chat'' mode, the default is not. 

\subsubsection{Additional Revenue Streams}
Verified creators on \mdf can accept donations, sell video {downloads} (versus online playback), and offer paid technical support or other resources through \emph{\mdf tokens}, a first-party digital currency where the lowest denomination of 100~tokens can be bought using \$1.99 worth of BTC, or directly through their Bitcoin address. 
After inspecting the \mdf token purchase process, we found that token purchases occur only through a third-party cryptocurrency payment processor. While Bitcoin transactions are public, we would only know what address(es) to inspect if we purchased tokens ourselves, directly contributing to site revenue. Further, without making repeated token purchases over time, we would not be able to determine if transactions are split across different addresses. Therefore, we chose to not pursue further analysis of \mdf token purchase volumes.
Though we cannot measure how many users pay to download video content, we observe that the advertised download price varied greatly, ranging from \$2~to \$32~in tokens. 

Similarly, we collected and attempted analysis for 292 unique Bitcoin addresses provided in designated video metadata fields for direct donations. However, because of cryptocurrency transaction anonymity, we could not determine which transactions were related to deepfake services. Bounding transactions by associated \mdf accounts’ creation timestamps, we did find that 20\% of addresses received at least \$\num{1000}. Ultimately, we note that these estimates are too loosely bounded to make verified claims about deepfake-related revenue for these addresses. 
In addition, video descriptions on the tube site commonly (\num{15039} of \NUMVIDEOS~videos, 35\%) included third-party links. About half (49\%) are to payment sites or subscription signups on both mainstream (e.g., PayPal, Ko-Fi, Patreon) and adult content (e.g., OnlyFans, FansMine, Fan-Topia) platforms. 

\section{Deepfake Video Creation}
\label{sec:pipeline}

In addition to serving deepfake video content and providing a marketplace for buyers and sellers, \mdf also hosts discussions about the \textit{creation} of deepfake videos. 
Prior work has established that while nonexperts can successfully create deepfakes
, the ability to create \emph{realistic} deepfakes (e.g., not suspicious to a human or software detector) requires deep expertise using complex, difficult-to-use software~\cite{mehta2023can}.
%
An active community has emerged on \mdf to facilitate this process, using the forum to discuss best practices, offer tips and suggestions on improving deepfake quality, and provide detailed guides.
In this section, we analyze forum discussion and the most popular guides for deepfake video production to document the requirements, challenges, and community solutions for the deepfake creation pipeline.

\subsection{Methodology} 

We specifically perform content analysis on the ``\textit{DeepFake Video Creation Tools}'' section of the forum, which is designated for members to discuss how to create deepfakes and also hosts \textit{guides and tutorials} for deepfake creation, authored by expert community members, forum staff, administrators, moderators, and verified creators. The section contained 56.9\% of all threads and 58.8\% of all posts on the platform. 
%
We began by manually examining the most active guide threads (by number of replies), as well as guides linked in a pinned ``quick start'' post for newcomers. We then analyzed other linked threads, excluding topics unrelated to sexual deepfake videos, which uncovered 25~guide and tutorial threads containing \num{1229}~posts. We annotated discussions for troubleshooting topics or resource requirements for deepfake creation, surfacing three major themes: (1) access to facial image data, (2) software proficiency, and (3) powerful hardware. For each, we discuss the challenges posed and how the \mdf community has developed corresponding solutions.

We attempted large-scale, automated analysis of forum posts to understand the deepfake video creation process through BERTopic topic modeling. However, we encountered several obstacles. First, the generated topic clusters did not provide insight beyond the existing structure of the forum sections (e.g., posts discussing facesets, tools for deepfake creation, etc.). Second, this approach fails to scale revealing thematic nuances within clusters, requiring manual annotation of topic clusters to effectively apply this method. Therefore, we ultimately pursued our manual approach.

\subsection{Facial Image Data}
\label{ssec:pipeline_facial_image_data}

High-quality facial image data is ``the most important part of making a good deepfake,'' according to moderators. Creating a realistic deepfake requires a significant volume of facial image data: users who seek to create deepfakes are recommended to assemble 3K--15K facial images for a single target. 
Beyond the quantity and resolution of images, deepfake quality depends on the range of different angles and facial expressions, as well as a match in the lighting, facial structure, and skin tone between the target and the original individual depicted in the destination video. Deepfake creators must then clean and process these images to extract faces from backgrounds, identify facial landmarks, and align faces.
Assembling new sets of quality facial data is laborious, and data-related discussion is highly prevalent, accounting for \num{1266} (20.2\%) of threads and \num{4269} (12.7\%) of posts in the forum sections we studied. 

\vspace{3pt}
\noindent
\textbf{Crowdsourced Facesets.}\quad
When facial image data is limited, users turn to the forum to request and share ``facesets,'' which are curated dumps of facial image data for a particular target. We find that facesets are often freely shared, as only two of the 711~threads that pertain to sharing facesets were labeled as paid solicitations. Of these threads, 81.6\% contained links to third-party file sharing sites (Appendix~\ref{table:fileshare}). In comparison, 326~threads request the creation of \textit{new} facesets to target an individual; however, only 39.6\% of these threads are replied to, and only 8\% are returned a link to a file sharing domain. Only three new faceset requests were paid offers. Experts' discussion and purchase offers of facesets suggest that the curation of high-quality facesets is labor-intensive and valuable:%
\quoteforum{
    I need someone to make me quality facesets\ldots maybe 3 or 4 a month. I'll pay you in crypto (enough to pay some if not all your rent, depending on where you live)
}

\noindent Anecdotally, we observe that buyers can purchase deepfakes at a lower cost if they are willing to provide their own faceset. We suspect that the lower success rate of finding willing community members to create new facesets for free is because of the requisite labor. 
In short, not only do community members crowdsource in the face of obstacles posed by limited facial image data, but many community members are willing to share existing data for free.

\subsection{Software}
\label{ssec:pipeline_proficiency}

Most deepfakes on \mdf are created using DeepFaceLab (DFL) or FaceSwap (FS). 
DFL is preferred and lauded by the community for its acceptance of ``not safe for work'' (NSFW) usage, technical capabilities, troubleshooting support, and lower system requirements:%
\quoteforum{%
    The FS dev team will not acknowledge any NSFW. The DFL dev has no issues with NSFW.%
}\vspace{-5pt}
\quoteforum{%
    DFL is far ahead in dealing with face obstructions. FS is working on it, but nothing has been released yet.%
}

\noindent
Both tools are free and open source but are complex to use. As a result, significant community effort is dedicated to streamlining and troubleshooting their usage, which we anecdotally find affects the adoption and popularity of deepfake software tools. Troubleshooting comprised \num{3137} (50.1\%) of threads in the sexual deepfake video-related sections, representing 3,945 posts (41.6\%) on the forum in aggregate. 
DFL is the subject of three pinned guides; the main DFL guide has the most views and replies (2M and 855, respectively), and the DFL repository creator and maintainer also appears to be an active member of the \mdf community, with an account adorned with a ``Developer'' badge.
Several other authors of the DFL preprint~\cite{perov2021dfl} appear to be active members, with roles including developer, staff, administrator, and verified creator. 

To further investigate the role of open source software (OSS) tools, we extracted 936~Github links from forum discussion, uncovering 141~repositories. For each repository, we manually identified the type of software (e.g., image/video creation, audio/voice tools, generic machine learning libraries, system monitoring tools, datasets, etc.), whether the repository was a fork of another repository, whether the tool is permissive of NSFW content, and if the repository has any associated academic papers. 35 repositories (25\%) are forks, of which 13~are forks of DFL, and nine are forks of Roop, a ``one-click face swap'' library. Beyond just Roop, 17 (12\%) repositories are a new class of tools that we term ``limited-data'' tools, which are designed to produce deepfakes without substantial data. The class has emerged in response to the obstacles posed by collecting sufficient quantities of facial image data, with some claiming to be able to generate deepfakes with a single facial image. 


\vspace{3pt}
\noindent
\textbf{Roop, Rope, and FaceSwap.}\quad Two ``limited-data'' tools stand out on \mdf: Roop/Rope, which have a moderator-written guide linked in the quick start, and FaceSwap, which is the subject of the eighth-most active thread on the forum. In terms of output fidelity, some members praise the tools' potential to produce high-quality deepfakes for simple videos, while others lament their inability to do so for complicated videos. This parallels the striation of the deepfake market into demands for two categories: a lower-tier for ``cheapfakes'' and a higher-tier for complex scenes that require deeper expertise and extensive image data. The former is handled by these limited-data tools, whose output quality suffers for more complex tasks; buyers of the latter must then turn to the marketplace to pay an expert for their services.

Roop is a deprecated OSS project that contains controls to censor potentially NSFW outputs\@. \textit{Rope} forks Roop, removing safety filters and enhancing usability for deepfake creation with a GUI\@.
Of the nine Roop forks, seven (including Rope) circumvent Roop's protections. Additionally, we find that \mdf members engineered NSFW workarounds to three of the nine repositories that self-identify as SFW (i.e., they have safeguards for NSFW outputs) limited-data OSS tools. Even with the best intentions from OSS maintainers of ``SFW deepfake generators,'' our results suggest these are easily bypassed by expert deepfake creators, who are willing to share their solutions with the larger community.

\vspace{3pt}
\noindent
\textbf{Research Tools.}\quad 
In our annotation of 141~linked Github repositories, we identified that 60~repositories (including forks) were directly linked to 43~academic papers; in aggregate, these open source research code repositories were referenced in the forum over 400~times. The associated papers range in publication date: the oldest is a 2001 paper that achieves color correction between source and destination images using statistical analysis~\cite{reinhard2001image}, and the six most recent papers were published in 2023, the year we collected our forum data~\cite{liu2023dfl,radford2023robust,bain2023whisperx,an2023pano,pan2023drag,yuan2023reliableswap}. Of these papers, CVPR is by far the most prevalent venue, accounting for 13 of the 43 papers (30.2\%); the remaining 30 papers are spread across 21~other venues, including well-known conferences ICCV, ECCV, NeurIPS, ICML, and SIGGRAPH\@. The frequent and up-to-date references to reputable academic publications demonstrates that the \mdf community follows cutting-edge research. 

For the community favorite, DFL (the Github repository was linked 203~times), the corresponding paper was initially published as an arXiv preprint in 2020~\cite{perov2021dfl} and later published in the peer-reviewed journal \text{Pattern Recognition} in 2023. The delta between the two versions largely focuses on potential abuses of DFL, which both versions claim is most commonly used by ``VFX artists.'' In addressing the ethical issues of advancing deepfake technology, the authors ``admit that deepfake techniques may affect public discourse quality and infringe upon the citizens’ portrait rights'' when weaponized for ``misinformation, manipulation, harassment, [or] persuasion.'' Despite this, the authors argue that suppressing the publication would not stop its development and believe that it was their ``responsibility to publish DeepFaceLab to the academia community formally''~\cite{liu2023dfl}. Neither version directly addresses the use or abuse of deepfake technology to generate sexual imagery. In August 2024, the repository for DFL was disabled by GitHub for terms of service (ToS) violations.

\subsection{Hardware}
\label{sssec:pipeline_gpu}

Adequate hardware is chiefly important for creating deepfakes. GPU requirements specified by community members ranged from 4--12~GB of VRAM with recommended GPUs from NVIDIA and AMD\@. When asked for entry level recommendations, an authoritative member (staff/moderator) recommended an ``RTX 3060 minimum or used 1080Ti/2080Ti, nothing else is worth anything for deepfakes (too slow or not enough VRAM), you want 11--12 minimum, that should be plenty until you become really good, then you can get something better with 16--24~GB of VRAM.''  
Hardware costs and electricity costs appear to be prohibitive for some creators:
\quoteforum{
    I wish I could buy my own GPU. It's too expensive to import to my country, but maybe one day :(
}
\vspace{-5pt}
\quoteforum{
    the electric cost of running gpus 24/7  will be too high
}

\vspace{3pt}
\noindent
\textbf{Cloud GPU Providers.}\quad In response to the steep hardware requirement, many creators rely on cloud GPU services. Based on our analysis of the links shared on \mdf, the most commonly discussed is Google Colab (a hosted version of Jupyter Notebook that provides users access to GPUs via a freemium mode), accounting for 92~links, followed by Paperspace (26) and AWS (10). As one community member in 2019 noted: ``Colab works so much better than my own PC\@. Do not bother trying to create fakes on a Radeon RX 580, what took me a day is done in an hour using Colab.''
However, as the community discusses, cloud GPU providers pose unique challenges, including the nondeterminism of provisioned GPUs, unexpected runtime disconnections, and threats to users' privacy (e.g., not accepting cryptocurrency payments, server-side scanning).

Furthermore, cloud GPU users are subject to provider ToS; for instance, in May 2022, Google banned the use of Colab for training deepfakes~\cite{google-colab-ban}, which immediately caused ripples in the community:
\quoteforum{
    We can't make free deepfakes on Colab because Google banned it. If you have a free account your runtime will be repeatedly disconnected. Eventually they might ban you from Colab.
}
\noindent These obstacles have led to more scattered community hardware recommendations for deepfake creators, such as buying their own dedicated GPUs, finding alternative cloud GPU providers, or engineering evasion tactics for Colab's deepfake ban. The arising cat-and-mouse dynamic between cloud GPU platforms and deepfake creators echoes what the security community has observed with other forms of online abuse, opening a space for ``bulletproof'' cloud GPU providers that are more permissive of abusive uses. However, the community support for these alternative providers is more sparse, reintroducing the challenges of software proficiency for nonexperts interested in deepfake creation (Section~\ref{ssec:pipeline_proficiency}). A staff member and moderator responded to a member asking about the tradeoffs between Colab and another provider, Runpod, in light of the Colab ban:
\quoteforum{%
    I actually never used Runpod, so maybe check out YouTube for tutorials. Or you can try Paperspace, but it might cost more. If electricity costs keep rising then I might have to use cloud service bullshit too.
}

\noindent Aligned with prior explorations of the efficacy of deplatforming abusive actors~\cite{chandrasekharan2017you,levchenko2011trajectories,kanich2008spamalytics,bellingcat2023any} and the pursuit of target hardening~\cite{redmiles2024friction}, we anecdotally find that while major platform policy changes do not exhaustively eradicate abuse, they do significantly increase the cost of deepfake creation; such changes in the deepfake value chain may be enough to turn away less capable (e.g., financially, technically) creators and stem the growth of this ecosystem.

\subsubsection{Google Colab Deepfake Ban}

In May 2022, Google banned using Colab notebooks for training models to generate deepfakes~\cite{google-colab-ban}. In the following month, \mdf members began to respond with reports of service interruptions, calls for help to the community, and speculations of how the ban was implemented and enforced:
\quoteforum{
    Maybe Google is detecting DFL usage then terminating those Colab notebooks. I hope it's temporary I don't know any other free options.
}

\noindent Members noted that Colab was flagging deepfake-related notebooks by detecting use of the DFL library. Almost immediately, the \mdf community sought to identify opportunities to circumvent the ban. Several members shared trivial yet effective solutions that obfuscated the source of deepfake code by forking the original DFL repository and then cloning the fork, reducing traceability to the banned DFL repository:
\quoteforum{
    You don't need to know how to code, just copy and paste the notebook cell changes. This simple fix should take you at most 30 minutes and lets you circumvent Google's detection of DFL.
}


\noindent While discussion of Colab has slowed since the ban, continued reports of service interruptions and iterative development of ban evasion solutions illustrate the cat-and-mouse game between Google and the community. The community concluded that Google's detection of DFL naively relies on a blocklist, resulting in recommendations to ``not share [their DFL] Github fork and try to give [it] a unique name.'' The risk of informing Google's deepfake detection has also led to some members seeking to move ban evasion support to private channels, asking others to ``please DM if [they] know of any tricks.'' As of July 2024, it is unclear whether community solutions still work today; some users report that they are able to use Colab, whereas others report errors. 

\subsection{Community Support}
\label{sec:community}

To support users in the complex process of creating deepfakes, \mdf provides a place to ask technical questions and receive help from the community.
To understand community norms, we analyze \num{3256}~\textit{question threads} (i.e., threads in a questions subsection) containing \num{14521}~posts from two ``\textit{Questions}'' subsections for parent sections ``\textit{Deepfake Video Creation Tools}'' and ``\textit{Stable Diffusion}.'' 
The \mdf community appears responsive and welcoming to newcomers. 80.6\% of question threads on \mdf received a reply, with the median time to receive a reply of 3.4 hours ($\mu=5.8$ days, $\sigma=39.3$ days), which is similar to Stack Overflow, where 80\% of posts receive a response and posts have a median response time of 40~minutes ($\mu=11.7$ days, $\sigma=99.2$ days). While \mdf' median time to response is longer than that of Stack Overflow, we suspect that this is largely a function of community size; as of August 2024, the Stack Overflow community had 26M~members~\cite{stackexchange2024sites}, as opposed to the 611K members of \mdf. 
On the other hand, the mean response time on \mdf is half that of Stack Overflow, illustrating that the distribution of response times on \mdf is much tighter. We also find that 72\% of questions on \mdf receive a response in less than three days. Therefore, while members of \mdf may not receive instantaneous responses, they can reliably expect to receive technical support within a reasonable time frame. Newcomers with accounts under 45~days old create 39.3\% of all question threads


Unlike Stack Overflow, which is known to enforce a ``no pleases or thank yous'' rule~\cite{stackexchange2024sites}, the \mdf community does not have any rules regarding the tone or delivery of questions in these forum sections. 
Anecdotally, we observe that the community encourages pleasantries and more personal connections between members, perhaps in part because enterprising members are looking to attract clientele:
\quoteforum{
    DM me for personal help, I offer private tutoring. Or you can read the beginners' guides I wrote for using DFL for deepfakes.
}


\noindent Additionally, active members of \mdf appear to pride themselves on building and maintaining authentic community, which may be a competitive advantage that allowed them to succeed over other deepfake sites:
\quoteforum{
    We are the only legit deepfake site with an actual community, the last thing we want is to scam our members.
}

\noindent In conclusion, the \mdf community provides fast technical support for deepfake creation, welcomes curious and inexperienced new members, and permits cordiality between its members. This inclusive environment likely helps the community to grow, exacerbating the threat and potential harms of abusive deepfakes.




\section{Discussion}
\label{sec:discussion}
Deepfake IBSA causes and amplifies a wide range of real harms. It is commonly used to financially sextort targets, generate ``revenge pornography,'' and explore sexual fantasies without obtaining consent~\cite{wei2024understanding,burkell2019nothing,kshetri2023economics,okolie2023ibsa}.
Development of generative AI technology has made deepfake capabilities more accessible, promoting widespread use---including by children~\cite{maiberg22clicks}.
In this work, we: (1) measured the growth of \mdf and the targets of deepfake videos; (2) characterized the participants of the deepfake market and analyzed the prices of bespoke sexual deepfakes; and (3) documented the required components of deepfake creation and how the community supports members in the process. We synthesize the implications of our findings on disincentivizing deepfake NCII production and mitigating its harms below. 

\subsection{\mdf as a Community}
\label{ssec:disc_community}
\mdf has grown immensely over the past few years, drawing over 1.5B~views of 43K~sexual deepfake videos as of November~2023. The scale of engagement highlights that consumption and discussion of sexual deepfakes has become mainstream and occurs openly online.
This growth has led to the maturation of its profit-driven marketplace for sexual deepfakes. As the market continues to develop, we expect seller specialization to evolve in tandem, similar to other specialized labor markets~\cite{sunwoong1989labor}. The market will likely also expand. For example, in response to tightening margins on celebrity deepfakes, one \mdf member suggested that you could make money by applying existing tools to do ``many things'' beyond celebrity fakes (Section~\ref{ssec:cost}).


\vspace{3pt}
\noindent
\textbf{Deepfakes as Art.}\quad
The \mdf community is welcoming and responsive; many of its members and leaders share a value system that positions deepfake creators as \textit{artists} and the sexual deepfakes they produce as \textit{artistic contributions} that benefit the community. This unique motivation differentiates deepfake creators from previously studied cybercrime and online abuse actors~\cite{thomas2021hatesok,chng2022hacker}. To better understand and disincentivize this class of perpetrator, we identify three areas for future work: first, characterizing deepfake creators' perceptions of the benefits and harms of their work; second, investigating the applicability of interventions for deepfake creators, such as targeting inaccurate meta-cognitive (i.e., second-order) beliefs and nudging, which prior work has found effective for combating misinformation~\cite{jeffrey2021understanding} and poor cybersecurity practices~\cite{zimmerman2021nudge}, respectively; third, drawing on Insoll et~al.'s study of child sexual abuse material offenders~\cite{insoll2024factors}, investigating the prevalence of help-seeking behavior (i.e., wishing to change) among deepfake creators to inform more effective interventions. 

\vspace{3pt}
\noindent
\textbf{Open Source and Academic Work.}\quad
Core contributors to \mdf are technically skilled and eager to share their expertise in creating deepfakes. 
Contrary to the beliefs of some open source developers~\cite{widder2022ethicalai}, denying technical support is not a viable intervention for deepfake tool abuse. 
Furthermore, significant sharing of academic work within the community shows that its members closely follow and utilize academic research, particularly from AI venues, highlighting the dual-use potential for many computing innovations. Similar to the OSS community, many instances of academic research have used technology neutrality and inevitability to justify the publication and dissemination of scientific developments. We acknowledge that the solution does not lie in silencing research findings or technological innovation, but we argue that researchers and developers should consider principles of ``target hardening'' in their own work to: (1) choose how to distribute their work, (2) rigorously evaluate and improve upon existing AI alignment methods, and, when possible, (3) preemptively design and integrate safeguards to promote the intended use and discourage the abuse of their tools.

\subsection{Deepfake Abuse Mitigation is Adversarial}
Technical solutions for preventing deepfake creation will likely be limited, as evidenced by the successful efforts of \mdf members to circumvent anti-abuse mechanisms added by cloud providers. While identifying platform-centered choke points (e.g., payment processors, hosting providers, ad providers) can reduce harm, as the security community has concluded for numerous other ecosystems~\cite{franklin2007inquiry,han2022infra,krebs2010body}, the emergence of alternative ``bulletproof'' service providers and cryptocurrencies has limited the efficacy of policy-centered mitigations like deplatforming. 
In the face of complex security problems with no tractable, perfect solutions, we emphasize the impact and importance of target hardening~\cite{redmiles2024friction}; we also argue that legislation may prove a promising approach for curbing the creation of nonconsensual sexual deepfakes.


\vspace{3pt}
\noindent
\textbf{Banning Deepfake NCII.}\quad
A \mdf staff member shared that the enactment of legislation internationally has already begun to complicate monetizing sexual deepfakes, implying that laws have sway over the deepfake economy (Section~\ref{ssec:cost}). In February 2024, following a push for federal legislation protecting against deepfake abuse~\cite{sarnoff2024taylor}, forum members asked staff for their reflections on the implications of a federal deepfake ban, to which a \mdf staff member replied:
\quoteforum{
    If deepfakes are banned, I could finally do something else with my life without feeling bad about quitting. I'll also have time for creative hobbies that I can publicly share instead of making porn
}

\noindent Laws may therefore dissuade even highly skilled, core members of the community. 
We argue that augmenting our existing mitigation approaches by incorporating legislative protections has potential to more effecitvely stem deepfake abuse.


\vspace{3pt}
\noindent
\textbf{Setting Ethical Norms for Deepfakes.}\quad
We also observed that \mdf' rules defining celebrities as acceptable targets of deepfake NCII were laxly enforced, indicating a difference between the community's written rules and normative behaviors in practice.
Echoing Timmerman et~al.'s findings~\cite{timmerman2023studying}, we presented evidence that members' beliefs regarding the ethics of deepfake NCII creation ranged widely (Section~\ref{ssec:consumers}).
Such dissonance even within the \mdf community illustrates the broader state of flux surrounding the ethics of deepfake content. This liminality presents an opportunity for policy impact beyond raising deepfake creation cost disincentives, through guiding education and legal process approaches. 
We encourage policymakers to update sexual education curriculum to confront technology-mediated sexual interactions; through early education regarding concepts such as sexual safety, consent, and bodily autonomy in digital-physical ecosystems, we can begin to shape resilient norms surrounding the creation and consumption of deepfake content that proactively mitigate its harms.

\vspace{3pt}
\noindent
\textbf{Seeking Justice and Reducing Harm.}\quad Finally, we preemptively broach legal considerations in light of the mainstreaming of deepfake content and increasing commoditization of deepfake creation tools, as evidenced by the growth of \mdf and its discussion of limited-data tools. We reiterate Pfefferkorn's recommendations regarding justice pathways for victims and perpetrators of deepfake NCII: that legislators consider the implications of criminal prosecution for preventing deepfake IBSA---especially regarding children. Due to rising popularity of commodity deepfake NCII tools among teens~\cite{maiberg24survey}, blanket criminal charges for perpetrators also affect children; Pfefferkorn argues that criminal prosecution of children is counterproductive, unjustifiable, and causes lasting harm~\cite{pfefferkorn2024teens}. Therefore, we argue that legislators carefully balance the trade-offs of retributive and restorative justice pathways and acknowledge the nuances of criminalization when children are also known perpetrators.

\section{Conclusion}
\label{sec:conclusion}
In this work, we analyzed the \mdf platform since its inception and characterized the community's motivations and deepfake creation process. Our investigation surfaces significant growth of deepfake NCII creation and consumption. Beyond nonconsensual sexual gratification and power as understood motivations for IBSA perpetrators, our qualitative analysis reveals novel motivations for buyers of deepfake NCII: the perception of deepfakes as art and the distribution of new custom deepfakes as a community contribution. We document the requirements for deepfake creation, the challenges it poses, and the respective community-driven solutions. Our work sheds light on the rapidly growing, complex ecosystem of deepfake NCII, and we highlight opportunities to design proactive and reactive interventions to better mitigate deepfake IBSA. 

\section*{Acknowledgments}
The authors thank Riana Pfefferkorn, Hans Hanley, and the anonymous reviewers for their feedback and suggestions. This work was supported in part by a Sloan Research Fellowship, a Magic Grant from the Brown Institute for Media Innovation, an NSF Graduate Research Fellowship \#DGE-1656518, and by the National Science Foundation under Grant Numbers \#2030859 and \#2127309 to the Computing Research Association for the CIFellows Project. Any opinions, findings, and conclusions or recommendations expressed in this material are those of the author(s) and do not necessarily reflect the views of the National Science Foundation.
\clearpage

\section*{Ethical Considerations}
\label{subsec:data_collection_ethics}

In this paper, we analyze \mdf, a public platform for buying, selling, and hosting sexual deepfakes. Our work is in the spirit of prior work by the security community~\cite{motoyama11analysis,tseng20tools,wei2024understanding,soska15measuring,gamage2022deepfakes}. However, our context is different, and we carefully evaluated the ethical implications of our specific research. We considered the potential harm to several stakeholders: the depicted and discussed targets, the \mdf platform, its users, and the researchers themselves. We discuss the ethical implications for each stakeholder below. 

\vspace{3pt}
\noindent
\textbf{Targets of Sexual Deepfakes.}\quad To avoid drawing undue attention to any targets mentioned in the forum, we redact any personally identifying information in the quotes we report and do not name any targeted individuals in the text.

\vspace{3pt}
\noindent
\textbf{The \mdf Platform.}\quad \mdf' terms of service prohibit using site content for commercial endeavors, redistributing content, and sending users unsolicited messages, but they do not otherwise prohibit scraping. The site's \texttt{robots.txt} only disallows crawling by the Internet Archive (\texttt{User-agent: ia\_archiver}). All scraped information is accessible to anyone with an account. (Account creation is free and does not require e-mail verification; site operators have also stated that users are not obligated to provide ``real'' information.) We acknowledge that in scraping the website, we increased server load and bandwidth on the part of the platform; however, we minimized the number of requests made to the platform, and our request volume is relatively low compared to the organic activity of tens of thousands of accounts that the platform services, such that our study has minimal impact on the platform.

We did not collect any video or image data, only relevant metadata required to answer our research questions. We considered whether to publish the name of the platform in this paper, as similar prior work opted to not reveal platform names as to not draw potential attention to their existence~\cite{tseng20tools}. However, we decided to name \mdf as it has been named directly in prior work~\cite{timmerman2023studying,popova2020reading} and is the top search result for celebrity sexual deepfakes. We hope that naming this platform can enable further research and policy on sexual deepfake creation and mainstreaming.

\vspace{3pt}
\noindent
\textbf{Users of \mdf.}\quad In examining public user profiles, the only potentially personally identifying information was the account username and site-assigned user ID. Accounts are pseudonymous: there are no fields for ``real'' names. We never attempted to deanonymize any user, and we did not interact with community members (e.g., via direct messages or public posts). In our report, we take care to exclude any user's account details when reporting quotes and opt to report only aggregate statistics to protect user privacy. Per a publication requirement from the USENIX Security program committee, we paraphrased quotes to mitigate the possible re-identification of quoted users.

\vspace{3pt}
\noindent
\textbf{Researchers.}\quad Given that this paper required qualitative analysis of text-based forum data, the researchers involved in these tasks were directly exposed to disturbing text discussion and deepfake content that could cause psychological or emotional distress.
The research team was aware of these potential harms but recognized that exposure to this content was unavoidable to gain domain knowledge, understand the affordances of rendered webpages, and verify collected data. We ultimately made an informed decision to move forward with our study because we believed in the timeliness and importance of investigating this ecosystem. To mitigate these harms, the researchers had regular check-ins regarding their psychological well-being, worked together as a team to process \mdf content, had access to therapy and counseling services, and took breaks as needed.

\section*{Data Sharing and Open Science}
We are committed to sharing our data and code with researchers at academic, non-commercial institutions seeking to conduct future work. 
We note that \mdf site content is available on the open Internet and that motivated actors can easily access the content. However, we do not want facilitate malicious abuse of \mdf data to potentially harm others.
Therefore, given the sensitive and disturbing nature of the data, rather than public distribution, we will selectively provide access to the resource \href{https://zenodo.org/records/14770467}{repository} upon manual verification of the requesting parties. We believe that making the data available to the research community with appropriate vetting (i.e., of affiliation, intended research purposes) balances adding friction for abuse with providing utility to the community.






{
\footnotesize
\bibliographystyle{plain}
\bibliography{paper}
}

\appendix
\label{appendix}


\section{\mdf Forum Section Map}
\setlist{nolistsep}
\label{app:forum_map}
The hierarchical map of the forum's sections.

\textbf{Announcements}
\begin{itemize}
    \item Important Announcements and News
    \item How to Use Features of MrDeepFakes.com
\end{itemize}

\textbf{Marketplace}
\begin{itemize}
    \item DeepFake Service Providers
    \item Hire a Freelancer
\end{itemize}

\textbf{Adult DeepFake Videos (NSFW / 18+)}
\begin{itemize}
    \item Verified Creator Celebrity Deepfake Megathreads
    \item Celebrity DeepFakes (Videos)
    \item Video Requests
    \begin{itemize}
        \item Paid Requests
    \end{itemize}
\end{itemize}

\textbf{Image Fakes (NSFW / 18+)}
\begin{itemize}
    \item Celebrity Photo Fakes
    \item Tools and Apps for Image Fakes
    \item Requests
    \begin{itemize}
        \item Paid Requests
    \end{itemize}
\end{itemize}

\textbf{Non-Adult Content (SFW / All Ages)}
\begin{itemize}
    \item SFW DeepFake Videos
    \item SFW Celebrity Photo Fakes
    \item Requests
    \begin{itemize}
        \item Paid Requests
    \end{itemize}
\end{itemize}

\textbf{DeepFake Video Creation Tools}
\begin{itemize}
    \item Guides and Tutorials
    \item Celebrity Facesets
    \begin{itemize}
        \item Requests
        \item Paid Requests
    \end{itemize}
    \item Pornstar Facesets
    \begin{itemize}
        \item Requests
        \item Paid Requests
    \end{itemize}
    \item Trained Models
    \begin{itemize}
        \item Requests
        \item Paid Requests
    \end{itemize}
    \item Celebrity to Pornstar Matches
    \item Tools and Apps for Deepfake Videos
    \begin{itemize}
        \item Unofficial Mods
    \end{itemize}
    \item Questions
\end{itemize}

\textbf{Stable Diffusion}
\begin{itemize}
    \item Guides and Tutorials
    \item Models
    \begin{itemize}
        \item Celebrity models
        \item NSFW (18+) models
        \item SFW models
    \end{itemize}
    \item Datasets
    \item Extensions and Tools
    \item Questions
\end{itemize}

\textbf{Lounge}
\begin{itemize}
    \item Introductions
    \item Discussion
    \item Off-Topic
    \item Claim Credit/Flag Videos
    \item Suggestions \& Feedback
\end{itemize}

\textbf{Other Languages}
\begin{itemize}
    \item Russian Community
    \begin{itemize}
        \item Discussion
        \item Guides and Tutorials
        \item Questions
    \end{itemize}
    \item French Community
    \begin{itemize}
        \item Discussion
        \item Guides and Tutorials
        \item Questions
    \end{itemize}
\end{itemize}

\section{File Sharing Platforms}
\label{table:fileshare}

\begin{table}[h!]
\small
\centering
    \begin{tabular}{l r}
        \toprule
        Domain                        &  \% Posts   \\          
        \midrule              
        mega.nz                       &     28.5        \\  
        drive.google.com              &     6.0        \\  
        zippyshare.com                &     2.8        \\  
        ibb.co                        &     0.9       \\     
        uploaded.net                  &     0.8       \\     
        mediafire.com                 &     0.6       \\     
        \bottomrule
    \end{tabular}
    \caption{\textbf{Top File Share Platforms}---The most prevalent file sharing service providers among the \num{2066} posts with file sharing links in faceset market threads.}
\end{table}

\section{Investigated Deepfake Production Guides}
\label{app:examined_guides}
The full list of the 25 guides and tutorials forum posts we examined for manual content analysis.

\begin{enumerate}
    \item \href{https://mrdeepfakes.com/forums/threads/guide-voice-cloning-app.7496}{https://mrdeepfakes.com/forums/threads/guide-voice-cloning-app.7496}
    \item \href{https://mrdeepfakes.com/forums/threads/guide-roop-rope-single-photo-deepfakes.14021}{https://mrdeepfakes.com/forums/threads/guide-roop-rope-single-photo-deepfakes.14021}
    \item \href{https://mrdeepfakes.com/forums/threads/guide-deepfacelab-2-0-google-colab-guide.1340}{https://mrdeepfakes.com/forums/threads/guide-deepfacelab-2-0-google-colab-guide.1340}
    \item \href{https://mrdeepfakes.com/forums/threads/discussion-optimizing-df-ud-models-training-by-reusing-techniques-workflows.18143/}{https://mrdeepfakes.com/forums/threads/discussion-optimizing-df-ud-models-training-by-reusing-techniques-workflows.18143/}
    \item \href{https://mrdeepfakes.com/forums/threads/guide-deepfacelab-2-0-guide.3886/}{https://mrdeepfakes.com/forums/threads/guide-deepfacelab-2-0-guide.3886/}
    \item \href{https://mrdeepfakes.com/forums/threads/guide-compositing-post-processing-video-editing-guides.3935/}{https://mrdeepfakes.com/forums/threads/guide-compositing-post-processing-video-editing-guides.3935/}
    \item \href{https://mrdeepfakes.com/forums/threads/guide-use-deepfacelab-on-runpod-desktop-with-the-linux-version.12394/}{https://mrdeepfakes.com/forums/threads/guide-use-deepfacelab-on-runpod-desktop-with-the-linux-version.12394/}
    \item \href{https://mrdeepfakes.com/forums/threads/guide-voice-cloning-with-talknet.9611/}{https://mrdeepfakes.com/forums/threads/guide-voice-cloning-with-talknet.9611/}
    \item \href{https://mrdeepfakes.com/forums/threads/guide-mve-machine-video-editor.6791/}{https://mrdeepfakes.com/forums/threads/guide-mve-machine-video-editor.6791/}
    \item \href{https://mrdeepfakes.com/forums/threads/guide-undervolting-and-underclocking-your-gpu-making-deepfaking-much-more-power-and-cost-efficient.10088/}{https://mrdeepfakes.com/forums/threads/guide-undervolting-and-underclocking-your-gpu-making-deepfaking-much-more-power-and-cost-efficient.10088/}
    \item \href{https://mrdeepfakes.com/forums/threads/guide-mve-editing-landmarks.8007/}{https://mrdeepfakes.com/forums/threads/guide-mve-editing-landmarks.8007/}
    \item \href{https://mrdeepfakes.com/forums/threads/rules-for-guides-and-tutorials-read-before-posting.9958/}{https://mrdeepfakes.com/forums/threads/rules-for-guides-and-tutorials-read-before-posting.9958/}
    \item \href{https://mrdeepfakes.com/forums/threads/quick-start-links-how-to-make-deepfakes-resources-links-guides-software-apps-websites.6284/}{https://mrdeepfakes.com/forums/threads/quick-start-links-how-to-make-deepfakes-resources-links-guides-software-apps-websites.6284/}
    \item \href{https://mrdeepfakes.com/forums/threads/guide-amp-training.15112}{https://mrdeepfakes.com/forums/threads/guide-amp-training.15112}
    \item \href{https://mrdeepfakes.com/forums/threads/guide-faceswap-how-to-get-started-part-three-conversion-and-faq.1145/}{https://mrdeepfakes.com/forums/threads/guide-faceswap-how-to-get-started-part-three-conversion-and-faq.1145/}
    \item \href{https://mrdeepfakes.com/forums/threads/guide-faceswap-how-to-get-started-part-one-requirements-and-installation.1089/}{https://mrdeepfakes.com/forums/threads/guide-faceswap-how-to-get-started-part-one-requirements-and-installation.1089/}
    \item \href{https://mrdeepfakes.com/forums/threads/guide-faceswap-how-to-get-started-part-two-extraction-and-training.1104/}{https://mrdeepfakes.com/forums/threads/guide-faceswap-how-to-get-started-part-two-extraction-and-training.1104/}
    \item \href{https://mrdeepfakes.com/forums/threads/rules-for-tools-and-apps-for-deepfake-videos-read-before-posting.9969/}{https://mrdeepfakes.com/forums/threads/rules-for-tools-and-apps-for-deepfake-videos-read-before-posting.9969/}
    \item \href{https://mrdeepfakes.com/forums/threads/rope-opal-easy-to-use-deepfake-tool-for-videos-win10-only-version-as-of-2024-02-19.16171}{https://mrdeepfakes.com/forums/threads/rope-opal-easy-to-use-deepfake-tool-for-videos-win10-only-version-as-of-2024-02-19.16171}
    \item \href{https://mrdeepfakes.com/forums/threads/swapface-org-the-best-tool-for-creating-stream-video-image-deepfake.11487}{https://mrdeepfakes.com/forums/threads/swapface-org-the-best-tool-for-creating-stream-video-image-deepfake.11487}
    \item \href{https://mrdeepfakes.com/forums/threads/facefusion-roop-continuation.16301}{https://mrdeepfakes.com/forums/threads/facefusion-roop-continuation.16301}
    \item \href{https://mrdeepfakes.com/forums/threads/best-deepfake-porn-website-for-video-faceswap.20858}{https://mrdeepfakes.com/forums/threads/best-deepfake-porn-website-for-video-faceswap.20858}
    \item \href{https://mrdeepfakes.com/forums/threads/download-tutorial-uncensor-japanese-adult-videos-jav.2186/}{https://mrdeepfakes.com/forums/threads/download-tutorial-uncensor-japanese-adult-videos-jav.2186/}
    \item \href{https://mrdeepfakes.com/forums/threads/deepswap-ai-info-free-trial.11173/}{https://mrdeepfakes.com/forums/threads/deepswap-ai-info-free-trial.11173/}
    \item \href{https://mrdeepfakes.com/forums/threads/new-face-swap-tool-online-faceswap-akool-com.11240/}{https://mrdeepfakes.com/forums/threads/new-face-swap-tool-online-faceswap-akool-com.11240/}
\end{enumerate}

\section{Sexual deepfake video discussion}
\label{app:discussion_breakdown}
The breakdown, by number of threads and posts, of forum discussion topics relating to sexual deepfake videos. The total number of threads is 6,259, comprising 33,496 posts in the \href{https://mrdeepfakes.com/forums/#adult-deepfake-videos-nsfw-18.8}{Adult DeepFake Videos (NSFW / 18+)} and \href{https://mrdeepfakes.com/forums/#deepfake-video-creation-tools.5}{DeepFake Video Creation Tools sections}.

\begin{enumerate}[label=(\alph*)]
    \item Sharing work: \href{https://mrdeepfakes.com/forums/forums/verified-creator-celebrity-deepfake-megathreads.64/}{Verified Creator Celebrity Deepfake Megathread} and \href{https://mrdeepfakes.com/forums/forums/celebrity-deepfakes-videos.9/}{Celebrity Deepfakes (Videos)}
    \begin{enumerate}[label=(\roman*)]
        \item 352 threads (5.62\% of threads in these sections)
        \item 2,844 posts (8.49\% of posts in these sections)
    \end{enumerate}
    \item Requesting work: \href{https://mrdeepfakes.com/forums/forums/video-requests.10/}{Video Requests} and \href{https://mrdeepfakes.com/forums/forums/paid-requests.30/}{Paid Requests}
    \begin{enumerate}[label=(\roman*)]
        \item 1,241 threads (19.83\%)
        \item 5,145 posts (15.36\%)
    \end{enumerate}
    \item Reference material: \href{https://mrdeepfakes.com/forums/forums/guides-and-tutorials.15/}{Guides and Tutorials}
    \begin{enumerate}[label=(\roman*)]
        \item 79 threads (1.26\%)
        \item 4,833 posts (14.43\%)
    \end{enumerate}
    \item Data: \href{https://mrdeepfakes.com/forums/forums/celebrity-facesets.17/}{Celebrity Facesets} and \href{https://mrdeepfakes.com/forums/forums/requests.35/}{Free} / \href{https://mrdeepfakes.com/forums/forums/paid-requests.59/}{Paid} requests, \href{https://mrdeepfakes.com/forums/forums/pornstar-facesets.22/}{Pornstar Facesets} and \href{https://mrdeepfakes.com/forums/forums/requests.60/}{Free} / \href{https://mrdeepfakes.com/forums/forums/paid-requests.62/}{Paid} requests, \href{https://mrdeepfakes.com/forums/forums/celebrity-to-pornstar-matches.26/}{Celebrity to Pornstar Matches}
    \begin{enumerate}[label=(\roman*)]
        \item 1,266 threads (20.23\%)
        \item 4,269 posts (12.74\%)
    \end{enumerate}
    \item Models: \href{https://mrdeepfakes.com/forums/forums/trained-models.34/}{Trained Models} and \href{https://mrdeepfakes.com/forums/forums/requests.61/}{Free} / \href{https://mrdeepfakes.com/forums/forums/paid-requests.63/}{Paid} requests
    \begin{enumerate}[label=(\roman*)]
        \item 147 threads (2.35\%)
        \item 1,204 posts (3.59\%)
    \end{enumerate}
    \item Tools: \href{https://mrdeepfakes.com/forums/forums/tools-and-apps-for-deepfake-videos.18/}{Tools and Apps} and \href{https://mrdeepfakes.com/forums/forums/unofficial-mods.41/}{Unofficial Mods}
    \begin{enumerate}[label=(\roman*)]
        \item 37 threads (0.59\%)
        \item 1,256 posts (3.75\%)
    \end{enumerate}
    \item Troubleshooting: \href{https://mrdeepfakes.com/forums/forums/questions.12/}{Questions}
    \begin{enumerate}[label=(\roman*)]
        \item 3,137 threads (50.12\%)
        \item 13,945 posts (41.63\%)
    \end{enumerate}
\end{enumerate}

\clearpage
\onecolumn
\section{Paid Request Codebook}
\label{app:codebook}
\begin{table*}[h!]
\small
\begin{tabular}{p{0.2\textwidth}p{0.7\textwidth}r}
\toprule
Code                                &   Meaning     &   \% \\
\midrule
pleasure    &    sexual satisfaction, fantasy, or preference   &   23.7     \\
bullying    &    repeated psychological harm done to victim    &    0    \\
circum\_consent    &    circumvent the need to obtain consent   &   0    \\
revenge    &    counteraction to cause harm to victim in return for an "injury"   &  0   \\
masculinity    &    proving masculinity to a sexually deviant network (Henry \& Flynn 2019)   &  0.8   \\
sextortion    &    forcing a victim to do (sexual) acts by threatening share their explicit images    &   0  \\
reputation    &    cause injury to social reputation of victim   &   0   \\
community    &    collective archive for a community to enjoy or benefit from   &    7.1   \\
private\_use    &    requester specifies they want this video for private use   &    1.9   \\
join\_reason    &    requester implies the request is what brought them to the forum   &     2.6   \\
fetish    &    fetish content as defined in DSM-5   &   1.1    \\
art    &    discussion of content as artistic or the deepfake creators as artists   &  2.6     \\
delusional    &    discussion that positions the requester/payer as entitled to sexualizing the target   &   0.8   \\
show\_skill    &    request framing as an opportunity for creator to display skill/prowess over others   &   0.8   \\
creation\_interest    &    requester expresses interest in learning how to create their own deepfakes   &    1.5   \\
sfw    &    requester wants a SFW video   &   1.5  \\
lf\_csam    &    requester is seeking CSAM    &   0.4  \\
lf\_model    &    requester is seeking a model and not the output video   &   1.1  \\
lf\_faceset    &    requester is seeking a faceset (image dataset)   &   0.4   \\
lf\_data\_labor    &    requester is seeking labor for making deepfake image data usable   &    0.4    \\
lf\_audio    &    requester is seeking audio fake (replacement of voice)   &    1.1    \\
batch\_request    &    requesting multiple videos (can imply a long chain of commissions)   &    4.9   \\
provides\_src    &    requester describes private source materials that they can provide   &    11.7    \\
closed    &    thread closed for unknown reason   &  1.5   \\
completed    &    thread closed because the request was fulfilled   &    0.8   \\
has\_price    &    includes some price figure (concrete) or rate   &    4.1    \\
pm    &    offers a private communication channel   &   15.0    \\
quality    &    discussion of desire of good/high-quality material   &   12.0   \\
professional    &    discussion of desire of professionalism in their chosen deepfake creator   &   0.4   \\
turnaround    &    discussion of a concrete turnaround time from creator   &    1.9   \\
ethics    &    requester discusses the ethical considerations in some capacity   &   1.5   \\
legality    &    requester discusses the legality in some capacity   &   0.8   \\
compliance    &    requester discusses the compliance with the community rules   &   0.8   \\
extra\_multi\_target    &    willing to pay extra for multiple targets   &   0.4   \\
extra\_male    &    willing to pay extra because target is male   &   0.4  \\
extra\_angle    &    willing to pay extra because the angle is challenging   &   0.4   \\
extra\_body    &    willing to pay extra or wait longer because of challenges posed by target body match   &   0.8     \\
target\_sw    &    specified target already produces sexual content   &   1.5  \\
target\_obfus    &    if the target's name is not specified anywhere in the request   &  10.9   \\
target\_personal    &    specified target is from the requester's personal life   &  2.3   \\
target\_self    &    specified target is the requester themself   &   0.8  \\
trust\_verified    &    requester is seeking a verified creator (designation of reputation on MDF)   &   0.4   \\
trust\_reviews    &    requester is seeking a creator that has a lot of reviews from prior commissions   &   0.4   \\
trust\_portfolio    &    requester is seeking portfolio of candidate creators' past work to vet them        &    2.6   \\
\bottomrule
\end{tabular}
\caption{\textbf{Codebook for Paid Deepfake Requests}---%
    Codebook for categorizing the different themes that emerged from our thematic analysis of a sample (N=263) of paid requests for deepfake videos.}
\label{tbl:interview_codes}
\end{table*}

\end{document}